\documentclass[reprint,
superscriptaddress,
 twocolumn,
 nofootinbib,
 nobibnotes,
 longbibliography,
 amsmath,amssymb,
 aps,
pra,
 floatfix,
 tightenlines,
 10pt
]{revtex4-1}
\usepackage{physics}
\usepackage{xcolor}
\usepackage{float}
\usepackage{graphicx}
\usepackage{subcaption}
\usepackage{ulem}
\usepackage{soul}
\usepackage{caption}
\usepackage[colorlinks=true,citecolor=blue,urlcolor=blue,linkcolor=blue]{hyperref}

\RequirePackage{stix}

\newcommand{\figref}{FIG. \ref}
\newcommand{\appendixref}{Appendix \ref}
\renewcommand{\eqref}[1]{{Eq.~(\ref{#1})}}

\newcommand{\ii}{\mathrm{i}}

\usepackage{bm}
\renewcommand{\Vec}[1]{\bm{#1}}

\begin{document}
\title{Ring solitons and soliton sacks in imbalanced fermionic systems}

\author{Mats Barkman}
\affiliation{Department of Physics, Royal Institute of Technology, SE-106 91 Stockholm, Sweden}
\author{Albert Samoilenka} 
\affiliation{Department of Physics, Royal Institute of Technology, SE-106 91 Stockholm, Sweden}
\author{Thomas Winyard} 
\affiliation{School of Mathematics, University of Leeds, Leeds LS2 9JT, United Kingdom}
\author {Egor Babaev}
\affiliation{Department of Physics, Royal Institute of Technology, SE-106 91 Stockholm, Sweden}

\begin{abstract}
We show that in superfluids with fermionic imbalance and uniform ground state, there are stable solitons.
These solutions are formed of radial density modulations resulting in nodal rings. We demonstrate that these solitons exhibit nontrivial soliton-soliton and soliton-vortex interactions and can form complicated bound states in the form of ``soliton sacks". In a phase-modulating (Fulde-Ferrell) background, we find different solitonic states, in the form of stable vortex-antivortex pairs.
\end{abstract}

\maketitle
 \section{Introduction}
Solitons have long been understood to have profound consequences for the physical properties of fermionic systems \cite{jackiw1976solitons,RevModPhys.60.781,goldstone1981fractional,yefsah2013heavy}. Recently  new methods were developed to create and observe solitons in superfluid ultracold atoms \cite{yefsah2013heavy}, opening up a route to explore new regimes and properties 
\cite{lutchyn2011spectroscopy,efimkin2015moving,ku2016cascade,ren2019solitons,dutta2017collective}.
We will focus on the existence of solitons in so-called imbalanced Bardeen-Cooper-Schrieffer (BCS) fermionic systems. Such superfluids exhibit pairing between fermions with different magnitudes of Fermi momenta. For example such pairing has been considered in the context of dense quark matter \cite{alford2001crystalline}, 
mixtures of different ultracold atoms
\cite{Zwierlein_2006,Radzihovsky2011,samoilenka2020synthetic,bulgac2008unitary,radzihovsky2010imbalanced,son2006phase}, and  superconductors \cite{Bianchi2003,Ironbased,Uji2018,Mayaffre2014,coniglio2011superconducting,norman1993existence,cho2011anisotropic,Matsuda2007,lortz2007calorimetric,singleton2000observation,martin2005evidence,Agterberg2014,barkman2019antichiral}.
When the effects of imbalance are strong, the ground state of such a system can spontaneously break translation symmetry, by inducing periodic modulation in the complex order parameter.
Two commonly considered inhomogeneous ground states are the Fulde-Ferrell (FF) state \cite{FuldeFerrell1964}, which exhibits purely phase modulation, and the Larkin-Ovchinnikov (LO) state \cite{LarkinOvchinnikov1964}, which consists of purely density modulation. These two are jointly referred to as FFLO states.
Situations where both phase and density modulate, have also been found \cite{barkman2019antichiral,samoilenka2020synthetic}. 
Conversely, if the imbalance is weak, the ground state remains uniform.
In this paper,  we will show that fermionic imbalance can nonetheless change the properties of the system, even when the ground state is uniform, through the existence of energetically stable solitonic excitations \cite{ThesisNote}.

\section{Ginzburg-Landau model}
We consider imbalanced systems in the weak-coupling limit, close to the tricritical point, i.e., where the uniform, FFLO and normal phases meet.
Near the tricritical point the amplitude of the order parameter and its gradients
are small, justifying a Ginzburg-Landau (GL) free-energy expansion, which
 has been derived from the microscopic theory for various imbalanced systems \mbox{\cite{Buzdin1997,Radzihovsky2011,agterberg2000ginzburg,barkman2019antichiral}}.
 Fermionic population imbalance diminishes the coefficient of the second-order gradient term, which leads to the required inclusion of higher-order gradient terms. We will use the  GL model, originally derived in \cite{Buzdin1997}, which has been shown to be sufficient when describing effects far from the boundary \cite{Barkman_PRL_surface_FFLO,Samoilenka_PRB_surface_FFLO}, where the free-energy density reads
 \begin{equation} \label{eq: Rescaled Ginzburg-Landau free energy}
\begin{aligned}
f = \alpha |\psi|^2 - 2|\psi|^4 + |\psi|^6 -c_1|\grad{\psi}|^2 + \frac{c_1}{2} |\laplacian{\psi}|^2   \\
+ c_2 |\psi|^2 |\grad{\psi}|^2 + \frac{c_2}{8} \Big( (\psi^* \grad{\psi})^2 + (\psi \grad{\psi^*})^2 \Big),
\end{aligned}
\end{equation}
written in dimensionless units, where the field $\psi = |\psi|e^{\ii\varphi}$ is the complex order parameter.
The model applies for systems with pairing between electrically neutral fermions or superconductors where the coupling to the vector potential is negligible.
The two parameters $c_1$ and $c_2$ are positive constants and, for systems with a two-dimensional Fermi surface, are given to be  $c_1 = 8/3$ and $c_2 = 16/3$. Therefore the model can be described by a single parameter $\alpha$, which depends on both the temperature $T$ and the population imbalance, fixed by the Zeeman splitting energy $h$.
The rescaled spatial coordinate is measured in units of the length scale $L_0 = \frac{\hbar v_F}{k_B T_c} \ell_0$, where $\ell_0 = \ell_0 \left( \frac{T}{T_c}, \frac{h}{k_B T_c} \right)$ is a dimensionless quantity that diverges as the tricritical point $(T_\ast,h_\ast)$ is approached. Here, $v_F$ denotes the Fermi velocity and $T_c$ denotes the critical temperature at no population imbalance.
Details of the rescaling are presented in \appendixref{appendix: rescaling GL functional}.

\begin{figure}
\centering
\includegraphics[width=0.5\textwidth]{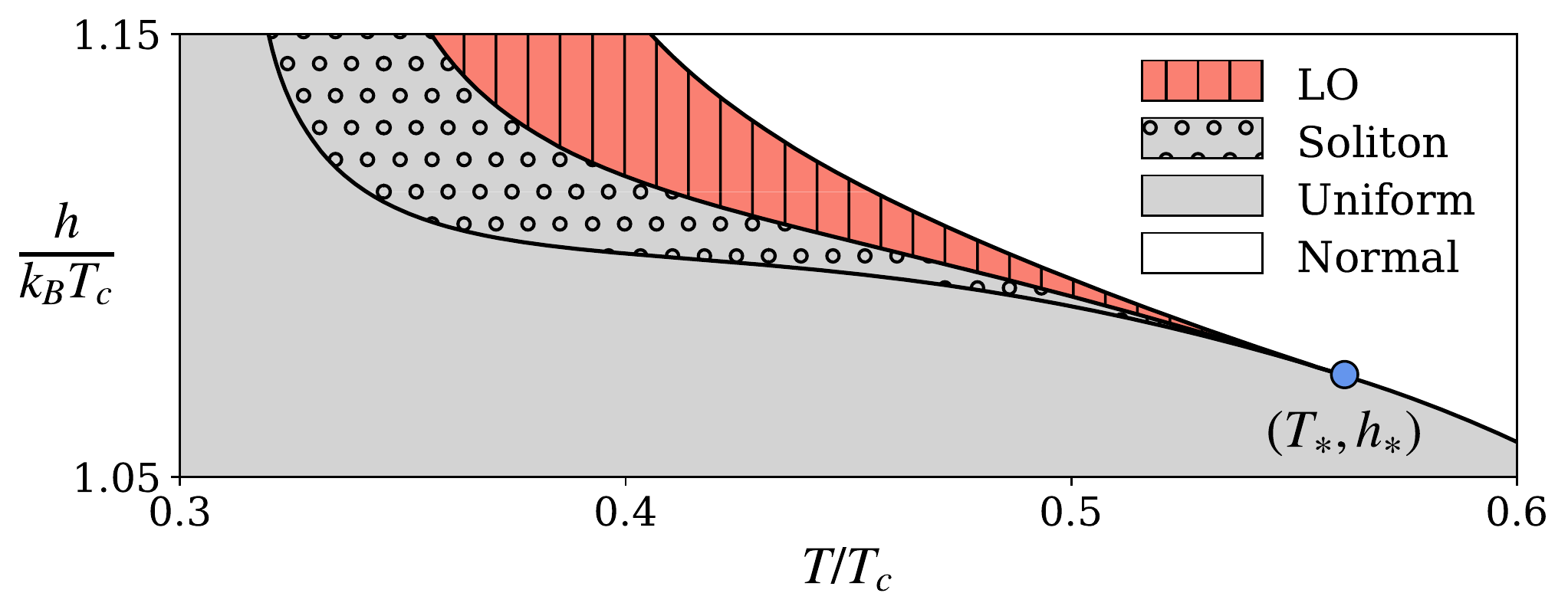}
\caption{
Phase diagram of an imbalanced fermionic system in the weak-coupling limit, close to the tricritical point $(T_*,h_*)$, where $h$ is the Zeeman splitting energy and $T$ is temperature. The  gray (red) background color indicates the uniform (LO) ground state. The regime with circular hatching denotes the regime in which there exist stable ring soliton excitations on top of the uniform ground state. The curve separating the uniform and soliton regimes is defined by $\alpha = \alpha_{c1} \simeq 0.56$, while the curve separating the LO and soliton regimes is defined by $\alpha = \alpha_c^{\rm{LO}} \simeq 0.857$.
} \label{fig: phase diagram}
\end{figure}

Since the second-order gradient term is strictly negative, there exists the possibility of nonuniform ground states. This can be seen in the two-dimensional model, where the configuration that minimizes the free energy transitions from a uniform state $\psi = \psi_{\rm{U}}$, where $|\psi_{\rm{U}}|^2 = \left( 2 + \sqrt{4-3\alpha} \right)/3$, to a density-modulating LO state at $\alpha^{\rm{LO}}_{c} \simeq 0.857$. This inhomogeneous state minimizes the free energy until the transition to the normal state at $\alpha = 4/3$. We note that the soliton solutions we find exist for the parameter region $0.56\lesssim \alpha < \alpha_c^{\rm{LO}}$, which is comparable to the size of the LO regime. 
These values of $\alpha$ can be translated into the corresponding values of the Zeeman splitting energy $h$ and the temperature $T$, as seen in \mbox{\figref{fig: phase diagram}}.

One feature to note regarding \eqref{eq: Rescaled Ginzburg-Landau free energy} is that it is mathematically related to the Swift-Hohenberg equation, which is also described by a fourth order partial differential equation. This model, initially formulated in the context of thermal convection \cite{swift1977hydrodynamic}, is commonly used to model pattern formation. 
The main differences between the GL model described by \eqref{eq: Rescaled Ginzburg-Landau free energy} and the standard Swift-Hohenberg equation are that the order parameter $\psi$ is complex valued and there is an additional coupling between the field strength and the gradient in the terms proportional to $c_2$ in \eqref{eq: Rescaled Ginzburg-Landau free energy}. A more detailed comparison between the models can be found in \appendixref{appendix: connection to SH}.

\section{Ring solitons}
We turn now to the numerical solutions of the free energy in \eqref{eq: Rescaled Ginzburg-Landau free energy}. We used a nonlinear conjugate gradient flow method both in finite element (FREEFEM \cite{FreeFem}) and finite difference schemes, which produced consistent results.

\begin{figure}
\centering
\includegraphics[width=0.45\textwidth]{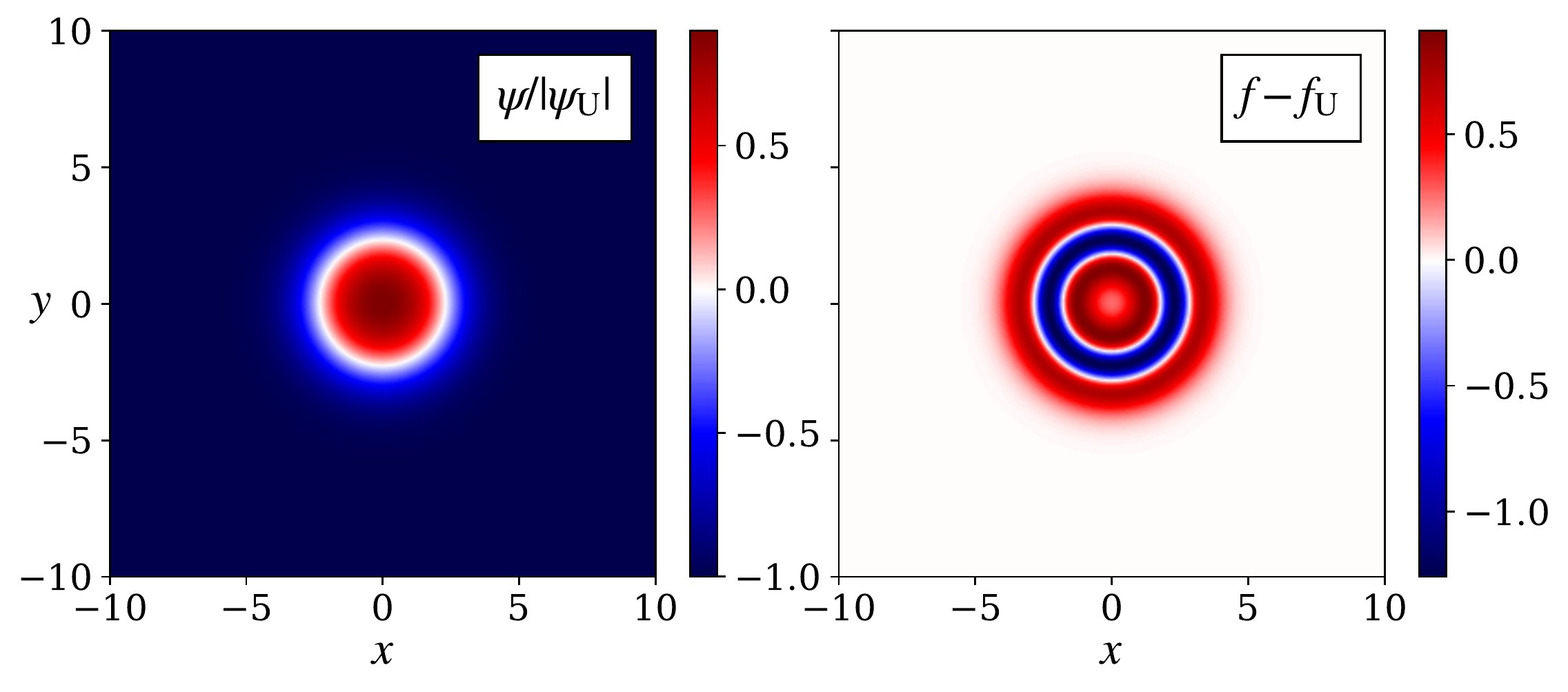}
\includegraphics[width=0.45\textwidth]{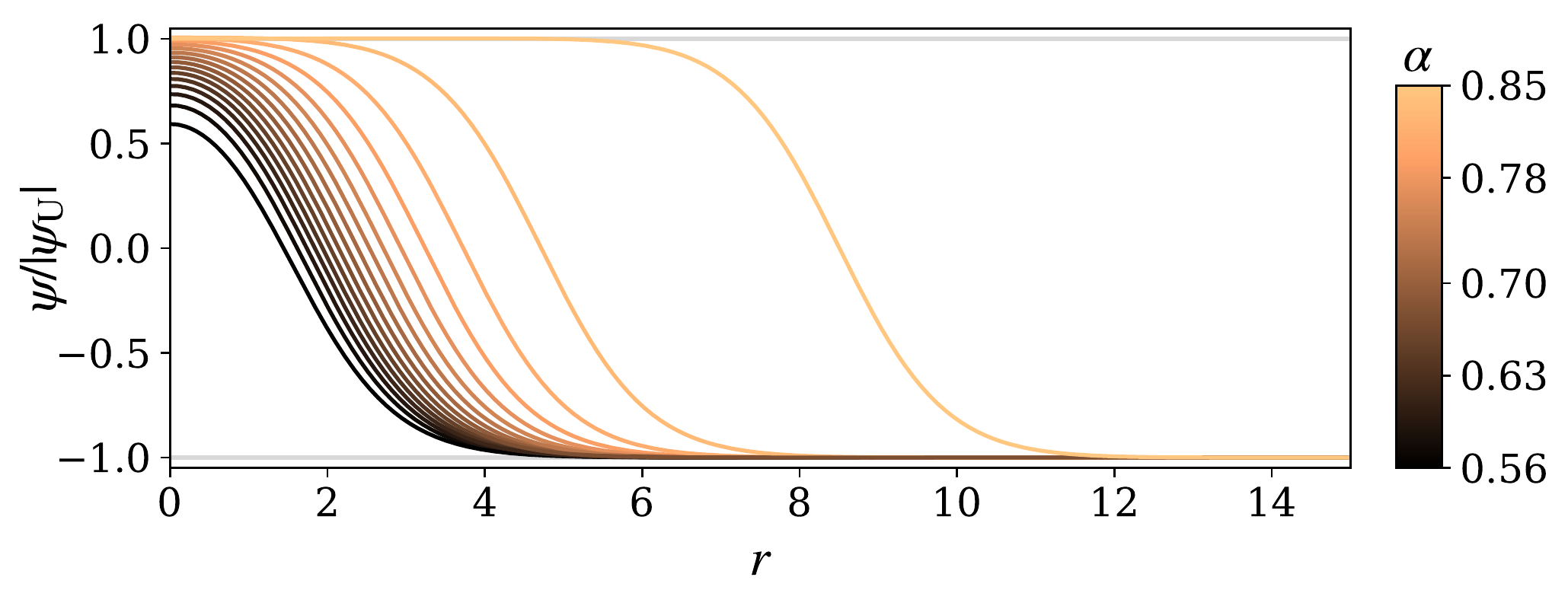}
\caption{The two upper panels show (left) the order parameter and (right)  the deviation of the free-energy density from the uniform ground state, for a ring soliton with one nodal ring ($N=1$) for $\alpha=0.7$ . The lower panel shows cross sections of the order parameter, for multiple values of $\alpha$,  of $N=1$ solutions, which were found to be stable for $\alpha \geq \alpha_{c1}\simeq 0.56$. The nodal radius increases with $\alpha$, such that for significantly large $\alpha$, the order parameter interpolates between the two ground state values $\pm |\psi_{\rm{U}}|$.} \label{fig: order 1 solitons}
\end{figure}

We find numerically that the two-dimensional GL free energy defined by the density in \eqref{eq: Rescaled Ginzburg-Landau free energy} has a number of local minima, in the form of solitons. The simplest subset of these soliton solutions retain the rotational 
spatial symmetry of the model, which we coin ``ring solitons''.
These radial solutions take the form  $\psi = g(r) e^{\ii\varphi}$, where $\varphi$ is a constant and $g(r)$ is a real profile function that modulates continuously between being positive and negative, before decaying to its ground-state value $\pm |\psi_{\rm{U}}|$ as $r \to \infty$. This leads us to characterise the solutions by the number of radial nodes ($g(r) = 0$) they exhibit ($N$). In the cases where $\varphi$ is constant, we assume without loss of generality that $\varphi = 0$ and thus $\psi$ is a real field. The $N=1$ solutions are displayed in \figref{fig: order 1 solitons}, where $\psi(r)$ changes sign once. It is important to note that the energy density deviation from the uniform state, plotted for $\alpha = 0.7$ in \figref{fig: order 1 solitons}, decays such that the total soliton energy is finite. 
This means that since entropy scales with system size, solitons will be thermally induced in the thermodynamic limit.

We found stable solutions for $\alpha \geq \alpha_{c1} \simeq 0.56$, which suggests that a system with sufficiently weak imbalance will not support these solitonic excitations 
(see \mbox{\figref{fig: phase diagram}} for the conversion into dimensionful parameters).
As $\alpha$ increases, the size of the soliton also increases, and the order parameter approaches $|\psi_{\rm{U}}|$ at the center of the soliton.

For higher values of $\alpha$ we find that $N > 1$ solutions become stable, first at $\alpha = \alpha_{c2} \simeq 0.733$ for $N=2$, followed by solutions with three and four nodal rings ($N=3,4$) at $\alpha_{c3}\simeq 0.784$ and $\alpha_{c4} \simeq 0.806$ respectively. The $N=1$-$4$ solutions are plotted in \figref{fig: profiles, excitation energy and radius} along with their energies and increasing nodal radii $R$ (i.e. $\psi(R)=0$). The $N=1$ solution has the lowest excitation energy above the uniform ground state. As the LO transition is approached ($\alpha \rightarrow \alpha_c^{\rm{LO}}$), the energies decrease, becoming zero relative to the uniform ground state at the transition. At $\alpha = \alpha_c^{\rm{LO}}$, a state that modulates between the values of the vacua indefinitely become stable, similar to the LO-modulating ground state. Therefore, despite only presenting solutions with four or fewer nodal rings, we expect that as the FFLO transition is approached, solutions with any number of concentric nodal rings become stable. Namely, for all natural numbers $N$, there exists an $\alpha_{cN}$ such that for $\alpha \in (\alpha_{cN},\alpha_c^{\rm{LO}})$, a solitonic excitation with $N$ concentric nodal rings is stable.

Surprisingly, the radial configurations of solitons that we found numerically can be approximated by a simple logistic function to reasonable accuracy. We first note that the LO state can be approximated as successive kinklike modulations. We can then approximate the transition value $\alpha_c^{\rm{LO}}$ to the LO state, by calculating when it is energetically favorable for a single kinklike modulation to appear $-$ namely when the total energy deviation from the constant ground state, of an infinite system, minimized with respect to $q$,  of $\psi = \psi_{\rm{U}} \tanh(q x)$ becomes zero. This was calculated to be  $\alpha \simeq 0.858$, which is remarkably close to the numerically computed value $\alpha_c^{\rm{LO}} \simeq 0.857$. This in turn leads us to approximate the $N = 1$ soliton in a similar way, by a radial kinklike profile 
\begin{equation}
\psi = \psi_{\rm{U}} \left( \tanh{q (r + R)}-\tanh{q (r - R)}  - 1 \right)    
\end{equation}
 If we then substitute this approximation into \eqref{eq: Rescaled Ginzburg-Landau free energy}, the total energy and nodal radius, when numerically minimized with respect to $q$ and $R$, are within an average of $1\%$ of the true numerical solution. Nonetheless, this approximation does not capture the asymptotics of the solution well.

\begin{figure}
    \centering
    \includegraphics[width=0.5\textwidth]{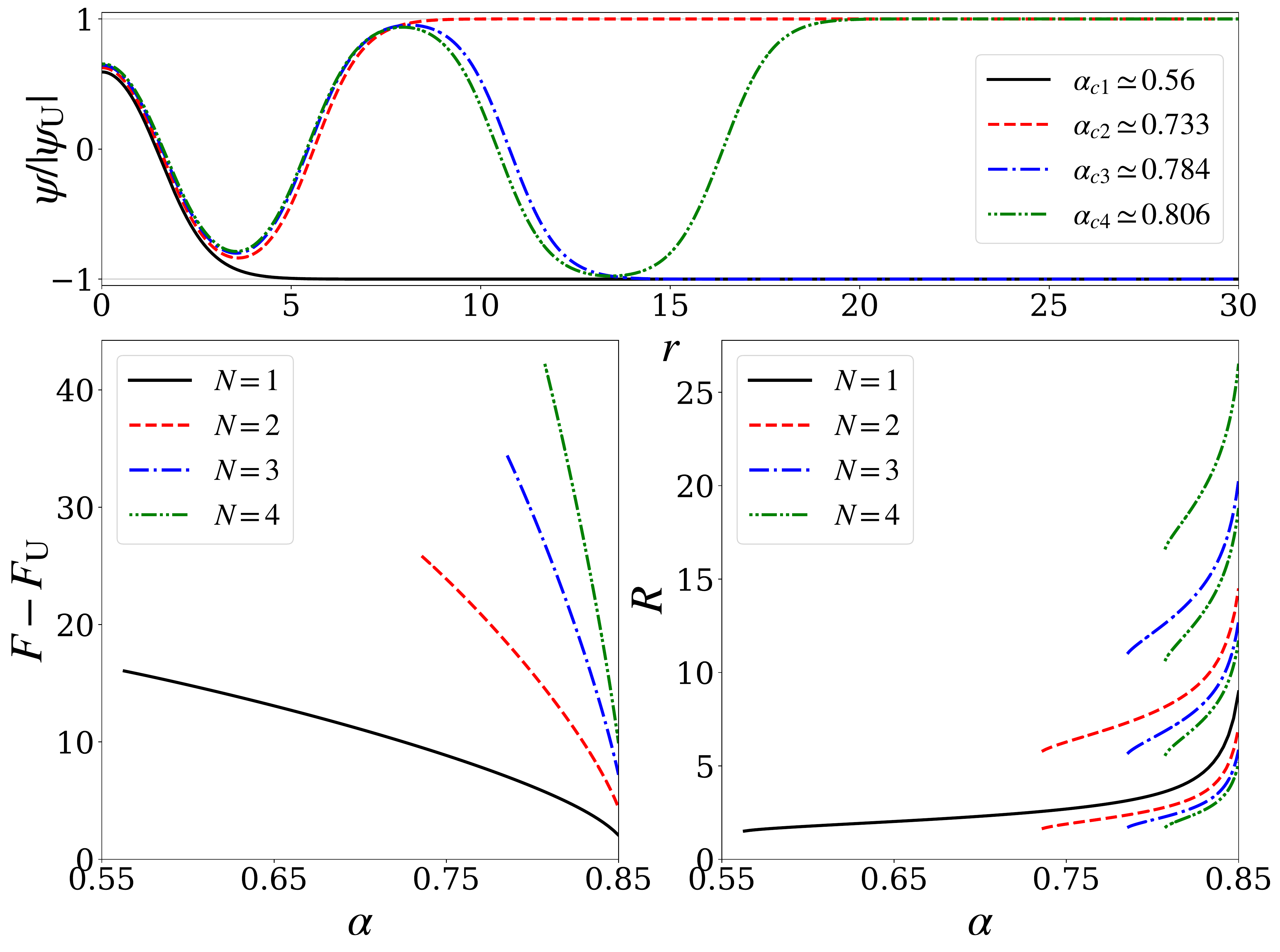}
    \caption{The upper panel shows radial solutions with $N$ concentric nodal circles, evaluated at the corresponding value of $\alpha$ at which they become stable ($\alpha_{cN}$). The associated excitation energy $F-F_{\rm{U}}$ (left) and nodal radii $R$ (right) of these solutions, as a function of $\alpha$, are shown in the lower panels. As $\alpha$ increases (the LO transition is approached), the excitation energy of the solution approaches zero and its nodal radii diverge.}\label{fig: profiles, excitation energy and radius}
\end{figure}

We can understand the stability and nature of these solitons by considering a crude approximation. Consider an $N=1$ soliton in the vicinity of the LO phase transition ($\alpha \lesssim \alpha_c^{\rm{LO}}$). Inspecting the numerical solutions in \figref{fig: order 1 solitons} suggests the approximation that $\psi \simeq \pm |\psi_{\rm{U}}|$ everywhere except for a small finite region centered on the nodal radius $R$.
We then approximate the various terms of the energy density as being independent of $R$, except for $\laplacian \psi \simeq \partial_r^2 \psi + \frac{\partial_r \psi}{R}$. We reiterate that this crude approximation is valid only when the nodal radius $R$ is large due to being close to the LO transition.
This gives the total excitation energy
\begin{equation}
F - F_{\rm{U}} \propto \int \left(A R + B + \frac{C}{R} \right) dr,
\end{equation}
where $A$, $B$, and $C$ depend on $\psi$. These terms then have the following physical interpretation:
$A_{\rm{tot}} = \int A dr$  corresponds to the energy per unit length of a straight nodal line, which in a uniform ground state is positive. Hence its contribution to the energy decreases as the radius $R$ becomes smaller.
This shrinking is balanced by the term $C_{\rm{tot}} = \int C dr$ (where $C \propto (\partial_r \psi)^2$) which represents the energy cost associated with increasing the curvature of the nodal ring $\propto \frac{1}{R}$.
These competing contributions lead to stable ring solitons with $R \simeq \sqrt{\frac{C_{\rm{tot}}}{A_{\rm{tot}}}}$.
Note, that $R \to \infty$ as $\alpha \to \alpha_c^{\rm{LO}}$, due to $A_{\rm{tot}} \to 0$ in this limit.
The argument above can be extended to $N > 1$, demonstrating that solitons with any number of nodal rings are expected to be stable if $\alpha$ is sufficiently close to $\alpha_c^{\rm{LO}}$.

We also considered the three-dimensional analog of the ring solitons with spherical nodal surfaces. However, despite evolving a number of initial conditions for different parameters, we did not find any stable solutions. This suggests that spherical nodal surfaces may be unstable in the considered regime.
However, we have not performed an exhaustive enough search to conclusively make this claim. It may still be possible that solutions with small energy barriers exist, requiring initial conditions close to the resulting configuration to relax to the local minima.

\section{Coherence lengths and long-range inter-soliton forces}

We can understand the long-range nature of the solutions by considering the linearized theory. This linearized theory will give the linear coherence lengths, determining the length scales at which the field recovers its ground-state value away from a perturbation. To that end we consider the field far from the soliton center, such that we can write it as a small perturbation $\varepsilon$ about its ground state 
\begin{equation}
\psi = \psi_{\rm{U}} + \varepsilon    
\end{equation}
 By assuming that any terms of order $O(\varepsilon^2)$ or higher are negligible, we acquire the tractable linearized equation,
\begin{equation}
    \nabla^4 \varepsilon + 2a \laplacian{\varepsilon} +b \varepsilon = 0,
    \label{eq:lin_eom}
\end{equation}
where $a = 1 - \frac{5 c_2}{4 c_1} |\psi_{\rm{U}}|^2$ and $b = 2(\alpha - 12 |\psi_{\rm{U}}|^2 + 15|\psi_{\rm{U}}|^4)$. As described in detail in \appendixref{appendix: linearization}, the solution to this linearized equation gives the asymptotic form of the field as
\begin{equation}\label{eq: asimptotics}
\varepsilon = \text{Re}\left[ C K_0(\mu r) \right],
\end{equation}
where $K_0$ is the zeroth-order modified Bessel function of the second kind and $C$ is a complex constant.
The coherence length $\xi = \mu^{-1}$ defines the length scale at which the deviation $\varepsilon$ decays. Importantly, in this imbalanced system, the coherence length is complex, since
\begin{equation}
     \frac{1}{\xi} \equiv\mu= \mu_R + \ii \mu_I = \sqrt{-a + \ii\sqrt{b - a^2}},
\end{equation}
where $\mu_R$ and $\mu_I$ are positive.
Therefore, in contrast to conventional superconductors and superfluids, the deviation $\epsilon$ oscillates while decaying.
At long range the behavior of the field is
\begin{equation}
    \varepsilon \to C_\infty \frac{e^{- \mu_R r}}{\sqrt{|\mu|r}} \cos{\left( \mu_I r + \phi_\infty \right)}\ \ \ \ \ \ \ \ \  \ (r \to \infty),
\end{equation}
where $C_\infty$ and $\phi_\infty$ are some real constants. Hence the tails of the solitons decay exponentially over the length scale $1 / \mu_R$, while their amplitude oscillates with period $2 \pi/\mu_I$. 
This effect is present in the states shown in \mbox{\figref{fig: order 1 solitons}}, but becomes visible only if the scales of the axes are changed.
Complex coherence lengths have previously been considered in other superconducting models \mbox{\cite{speight2019chiral}}.

\begin{figure}
    \centering
    \includegraphics[width=0.5\textwidth]{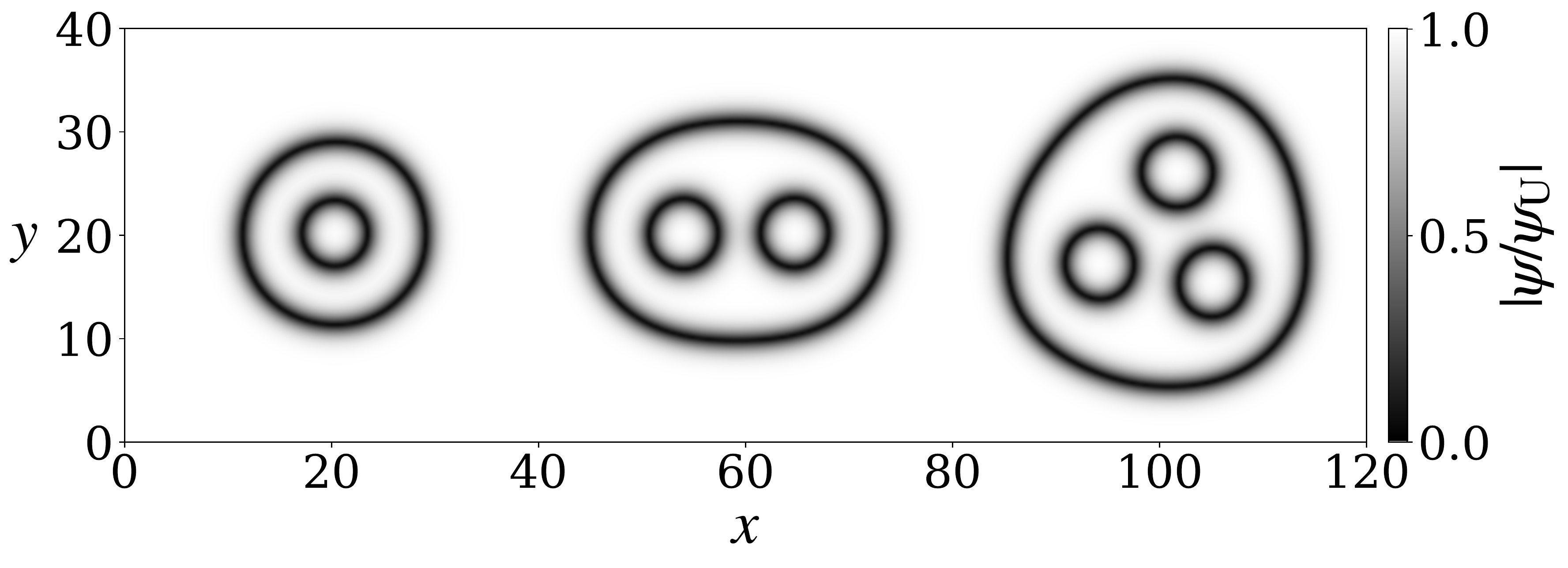}
    \caption{
    The state to the left is the $N=2$ soliton, where the simplest soliton $(N=1)$ is surrounded by an additional nodal ring. The middle and right states are composite objects, where multiple $N=1$ solitons are confined by an outer nodal ring. We call these composite structures soliton sacks.
    All solutions above were computed at $\alpha=0.825$.} \label{fig: soliton sacks}
\end{figure}

As we understand the asymptotic field behavior of our solitons, we can approximate the long-range intersoliton forces. We do this by considering two point sources that replicate the asymptotic fields of the interacting solitons and calculating the interaction energy between them \cite{Speight_vortex_forces}.
The total deviation from the ground state is assumed to be the superposition of the two asymptotic fields (given in \eqref{eq: asimptotics}) with constants $C^{(1)}$ and $C^{(2)}$, located at $\Vec{x}_1$ and $\Vec{x}_2$, respectively, where the distance $|\Vec{x}_1-\Vec{x}_2|$ is large. Details of the calculation are given in \appendixref{appendix: interaction}. This gives the interaction energy
\begin{equation}
    F_{\rm{int}} = - 2 \pi c_1 \sqrt{b - a^2} \text{Im}\Big( C^{(1)} C^{(2)} K_0( \mu|\Vec{x}_1-\Vec{x}_2| ) \Big),
\end{equation}
which is an oscillating function of separation distance. This predicts that there will be weakly bound states with period ${2 \pi}/{\mu_I}$ for solitons at large distances.

It is interesting to compare the radial solitons reported here with the related, yet distinct, radial solutions to the stationary Swift-Hohenberg (SH) equation \cite{SAKAGUCHI1996274,Lloyd_2009}.
The similarity is that both solutions exhibit stable radial oscillations.
Apart from the previously mentioned differences (complex order parameter and the additional terms proportional to $c_2$ in \eqref{eq: Rescaled Ginzburg-Landau free energy}),
 there is a key difference in that the ground state in \cite{SAKAGUCHI1996274,Lloyd_2009} is the homogeneous solution $\psi=0$, around which the solutions oscillate. In contrast, the nonlinear part of our solution modulates between two antipodal points on the $U(1)$ ground-state manifold.  At large distances, both the SH and our solutions decay exponentially, exhibiting oscillatory tails.

\section{Soliton Sacks}

In previous sections we considered only radially symmetric solutions, however it transpires that these represent only a small fraction of the solitonic solutions in the model described by \eqref{eq: Rescaled Ginzburg-Landau free energy}. As the LO transition is approached, we find more structurally complicated stable solutions that break rotational symmetry. Examples of these symmetry-breaking solutions are plotted in \figref{fig: soliton sacks}. These can be interpreted as soliton sacks, where a larger soliton confines a group of smaller solitons. This confinement is a completely nonlinear effect and cannot be explained by the asymptotic intersoliton forces. Such solutions are reminiscent of the ostensibly unrelated Skyrmion sack or bag solutions, which attract substantial interest in superconductors \cite{garaud2011topological,garaud2013chiral}, chiral magnets \cite{Rybakov_Skyrmion_sacks}, and liquid crystals \cite{Foster2019}.

\section{Soliton-vortex composite}

\begin{figure}
\centering
\includegraphics[width=0.45\textwidth]{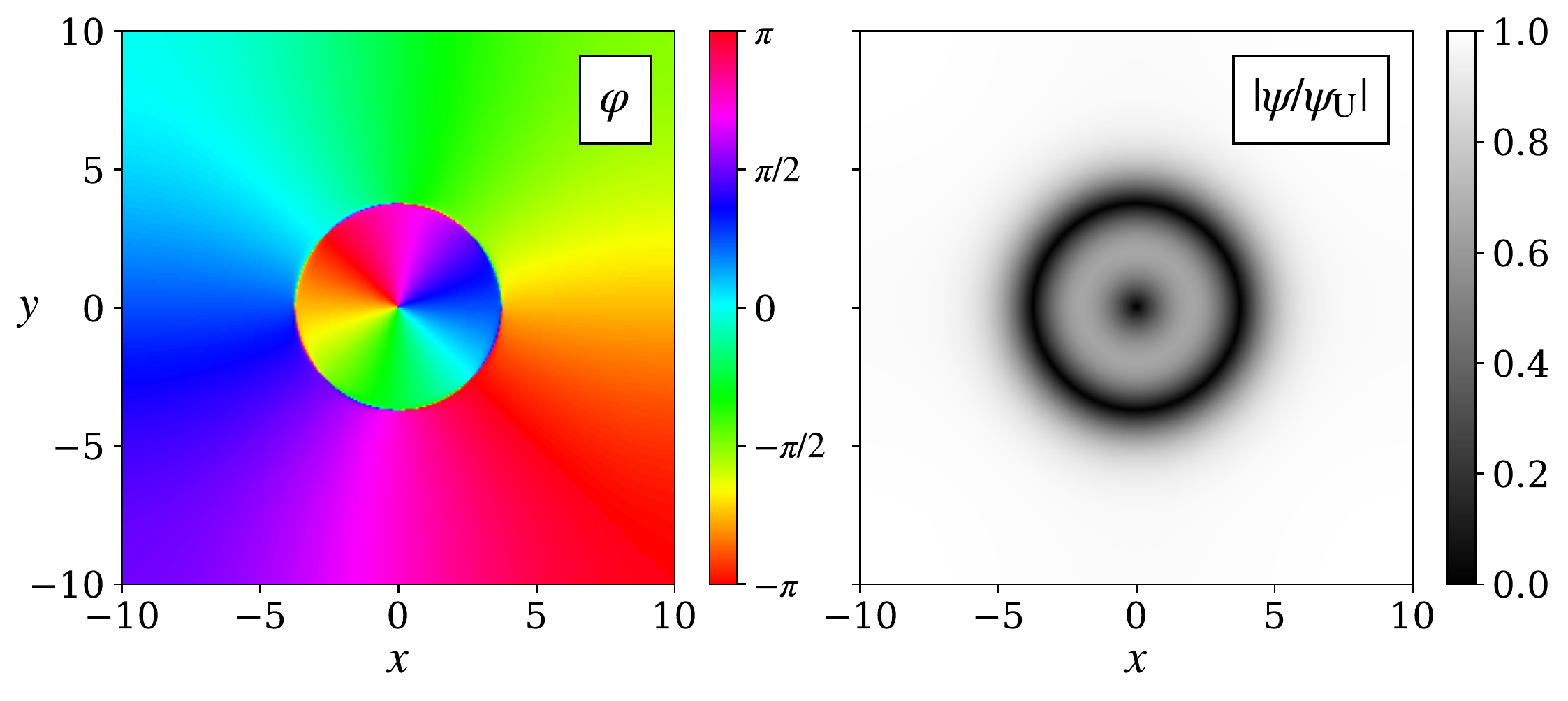}
\caption{A bound state of a vortex and a soliton, at $\alpha=0.7$. At the nodal line of the soliton the vorticity receives a $\pi$ phase shift. The energy of this composite topological defect is smaller than the energy of an infinitely separated vortex and soliton. } \label{fig: vortex+soliton}
\end{figure}

While we have demonstrated a rich spectrum of new solutions in imbalanced systems, the natural question is how they interact with the familiar soliton excitations, namely vortices. We consider a regime away from the LO instability (exemplified by the choice $\alpha=0.7$) where ordinary vortex solutions exist (namely, away from the regime where a vortex core induces an FFLO state as reported in \cite{inotani2020radial}). We find that the solitons and vortices form bound states, shown in \figref{fig: vortex+soliton}. The energy of this bound state is lower than the combined energy of a separate single vortex and soliton, but the energy of the bound state is larger than the energy of a single ordinary vortex. 

\section{Solitons in Fulde-Ferrell State: stable vortex-antivortex pairs}
\begin{figure}
\centering
\includegraphics[width=0.5\textwidth]{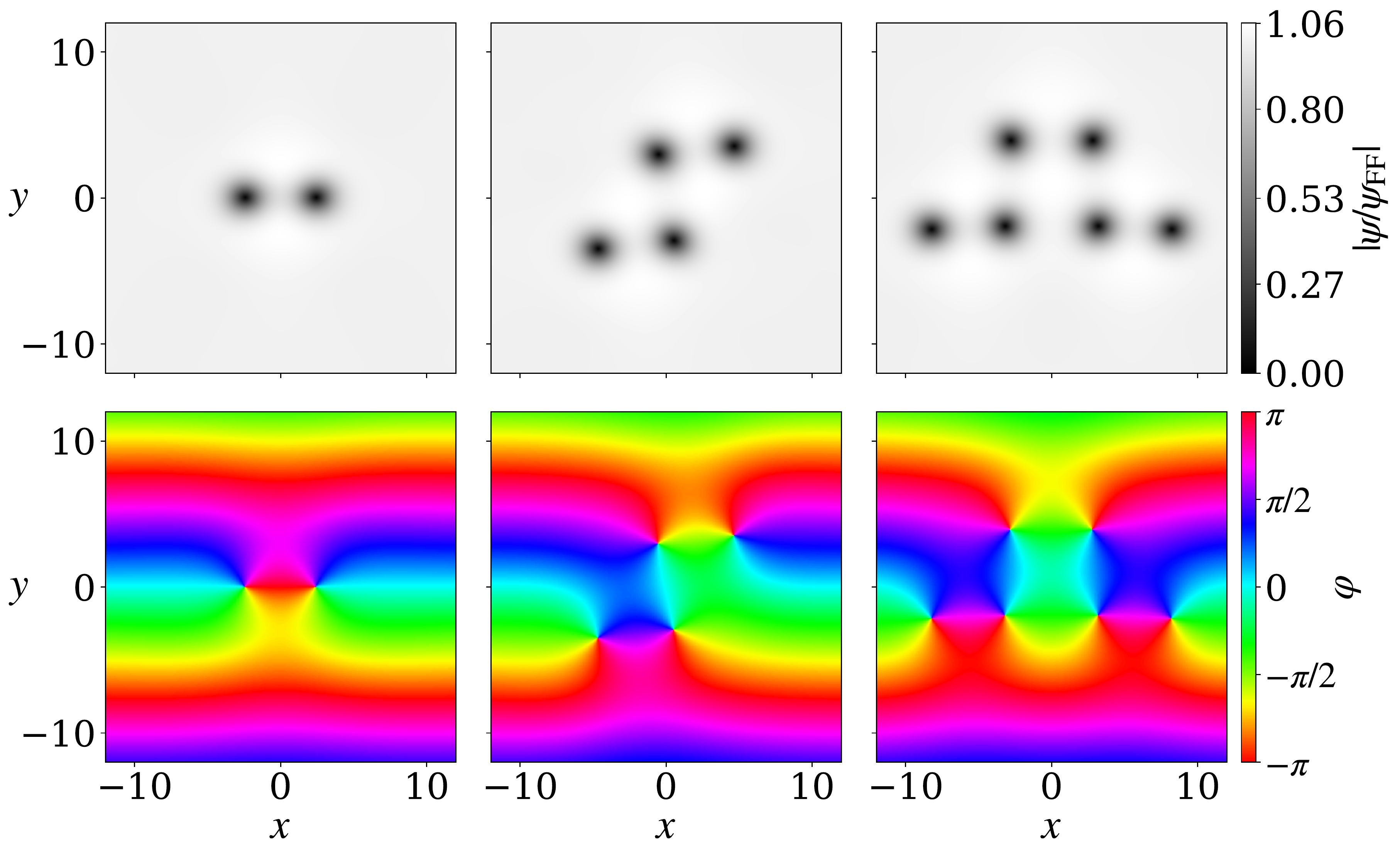}
\caption{
Solitonic excitations in FF state: stable vortex-antivortex pairs.
The pairs have long-range attractive interaction like in an ordinary superfluid, but in the FF background are protected from decay by a potential barrier. The solutions are shown for the GL model in {\eqref{eq: Rescaled Ginzburg-Landau free energy}} with $\alpha=0.5$ and $c_1=c_2=2$.} \label{fig: VAV pairs}
\end{figure}

Finally, our results prompt the question of whether or not these solitons exist over the background of another imbalanced state with uniform density: namely the Fulde-Ferrell (FF) state, where the background phase modulates. An example of a microscopically derived Ginzburg-Landau model for the FF state can be found in \cite{agterberg2000ginzburg}.
To model the FF state, without fine-tuning, we chose parameters $\alpha=0.5$, $c_1=2$, and $c_2=2$ phenomenologically. Our numerical studies suggested that the solitons described above, are not stable on top of the FF ground state $\psi_{\rm{FF}} = |\psi_{\rm{FF}}|e^{\ii qy}$ but that the Fulde-Ferrell state has its own stable solitonic excitations of a different kind: vortex-antivortex pairs. In contrast, we did not observe such a solution outside of the FF  regime. In \figref{fig: VAV pairs} we show a stable vortex-antivortex pair and examples of the structures that can be formed by multiple pairs.

\section{Conclusions}
In conclusion, we have shown that in BCS superfluids and superconductors, fermionic imbalance leads to solitonic excitations in the form of ring solitons. These solitons constitute a number of local minima of the free energy landscape. The solutions we find are related to, but distinctly different from, solutions of the Swift-Hohenberg equation. We have shown that these solitons have nontrivial nonlinear bound states: soliton sacks. 
Additionally, the long-range intersoliton forces predict bound states at larger separation, and the introduction of phase winding generates bound states of solitons with vortices.
We have also demonstrated the existence of stable solitonic states in a phase-modulating background, in the form of vortex-antivortex pairs.

In ultracold atoms, ring solitons could be created and observed by imprinting methods and standard density-sensitive techniques, due to the fact that order parameter modulation is typically accompanied by density modulation \cite{Samoilenka_PRB_surface_FFLO,Patton2020}. In a superconductor, such solitons could be observed via scanning tunneling microscopy.

Let us now remark on the particular case of the $N=1$ soliton, that was reported earlier by one of the authors \cite{ThesisNote}. Objects structurally similar to the simplest $N=1$ ring soliton have since been discussed independently in the study of unitary Fermi gases at low temperature \mbox{\cite{Magierski2019,Tuzemen2020}},  suggesting that radial solitons may appear under different circumstances from those in the BCS limit. It would be interesting to consider whether unitary systems support more complex bound states, such as soliton sacks, similar to those found in this paper.

Another avenue for further work is to study the dynamics of the presented static solutions. Introducing simple time dependence to the system will lead to dissipative dynamics, equivalent to gradient flow. Since the static solutions we present are local minima of the free-energy landscape, various initial conditions would simply relax to the static solutions presented. However, one interesting question is the behavior of the solitons in a dynamically driven system.
A further important question is the height of the energy barrier that prevents the solitons from collapsing. To determine the barrier height quantitatively, techniques such as the string method {\mbox{\cite{stringm,simplified_string_method}}}, recently generalized to superconducting models {\mbox{\cite{benfenati2020vortex}}}, should be used.

\section*{acknowledgments}
We thank Andrea Benfenati and Martin Speight for useful discussions.
The work was supported by Swedish Research Council Grants No. 642-2013-7837, 2016-06122, and 2018-03659, the G\"{o}ran Gustafsson Foundation for Research in Natural Sciences and Medicine, Olle Engkvists Stiftelse, and the UK Engineering and Physical Sciences Research Council through Grant No. EP/P024688/1. The computations were performed on resources provided by the Swedish National Infrastructure for Computing (SNIC) at the National Supercomputer Center in Link\"{o}ping, Sweden, partially funded by the Swedish Research Council through Grant Agreement No. 2018-05973.

\appendix

\section{Rescaling of Ginzburg-Landau functional} \label{appendix: rescaling GL functional}

In this paper we use the Ginzburg-Landau (GL) free energy expansion  derived in \cite{Buzdin1997}, starting from the following microscopic Hamiltonian for a spin-imbalanced superfluid
\begin{equation}
\begin{aligned}
    H = \int d^d{x} \Bigg\lbrace   \sum_{\sigma=\pm 1}  \Phi_\sigma^\dagger (\Vec{x}) \left[\frac{-\laplacian}{2m}  + \sigma h \right]\Phi_\sigma (\Vec{x})  \\ + \Big( \Delta (\Vec{x}) \Phi_{+1}^\dagger (\Vec{x}) \Phi_{-1}^\dagger (\Vec{x}) + \text{H.c.} \Big) \Bigg\rbrace,
\end{aligned}
\end{equation}
where $\Phi_\sigma $ is the fermionic quantum field operator, $h$ is the Zeeman splitting energy, $\Delta$ is the superfluid order parameter and H.c. denotes Hermitian conjugation. The resulting GL free energy functional reads
\begin{align}
       F  = & \int d^d x \Big\lbrace \alpha |\Delta|^2 + \gamma |\Delta|^4 + \nu |\Delta|^6 + \beta |\grad{\Delta}|^2 + \delta |\laplacian{\Delta}|^2 \nonumber \\ 
       & + \mu |\Delta|^2 |\grad{\Delta}|^2 + \frac{\mu}{8} \Big( \big(\Delta^* \grad{\Delta} \big)^2 + \big(\Delta \grad{\Delta}^* \big)^2 \Big)  \Big\rbrace,
\end{align}
where the coefficients $\alpha, \beta, \gamma, \delta, \mu$, and $\nu$ are functions of the temperature $T$ and the Zeeman splitting energy $h$. In natural units $\hbar = k_B = 1$, the coefficients $\alpha, \gamma$, and $\nu$ are given by
\begin{align}
    \alpha  & = -\pi N(0) \left( \frac{1}{\pi} \ln\frac{T_c}{T} + K_1(h,T) - K_1(0,T_c)\right), \\
\gamma   & = \frac{\pi N(0) K_3 (h,T)}{4},  \\
\nu  & = -\frac{\pi N(0)K_5(h,T)}{8}, 
\end{align}
where $N(0)$ is the electron density of states at the Fermi surface, $T_c$ is the critical temperature at $h=0$, and
\begin{equation}
K_n (h,T)= \frac{2T}{(2 \pi T)^n}  \frac{(-1)^n}{(n-1)!} \text{Re} \left[  \Psi ^{(n-1)}(z)\right],
\end{equation}
where $z = \frac{1}{2}- \ii \frac{h}{2 \pi T}$ and $ \Psi ^{(n)}$ is the polygamma function of order $n$. The remaining coefficients are given in terms of $\gamma$ and $\nu$ as $\beta = \hat{\beta} v_F ^2 \gamma$, $\delta = \hat{\delta} v_F^4 \nu$, and $\mu = \hat{\mu} v_F^2 \nu$, where $v_F$ is the Fermi velocity and $\hat{\beta}, \hat{\delta}$, and $\hat{\mu}$ are positive constants that depend on the dimensionality $d$ of the Fermi surface. The numerical values of $\hat{\beta}, \hat{\delta}$, and $\hat{\mu}$ in one, two and three dimensions are given in TABLE \ref{tab: Microscopic coefficients}. The possibility of inhomogeneous ground states arises in the parameter regime in which the gradient coefficient $\beta$ is negative. Since $\beta$ shares sign with the quartic coefficient $\gamma$, the inclusion of positive higher-order terms, both in density and momentum, is necessary.

For convenience we perform the following rescaling
\begin{equation}
    \psi = \frac{\Delta}{|\Delta_0|}, \quad \tilde{\alpha} = \frac{\alpha}{\alpha_0}, \quad \tilde{x} = q_0 x, \quad F = \frac{\alpha_0 |\Delta_0|^2}{q_0^d} \tilde{F},
\end{equation}
where $|\Delta_0|^2 = \frac{-\gamma}{2\nu}$, $\alpha_0 = \frac{\gamma^2}{4\nu}$, and $q_0^2 = \frac{-\beta}{2\delta}$ and the rescaled free energy $\tilde{F}$ reads
\begin{align}
   \tilde{F} & = \int d^d \tilde{x} \Big\lbrace \tilde{\alpha} |\psi|^2  - 2 |\psi|^4 +  |\psi|^6- c_1 |\tilde{\grad} \psi|^2 + \frac{c_1}{2} |\tilde{\nabla}^2 \psi|^2 \nonumber \\ & + c_2 |\psi|^2 |\tilde{\grad} \psi|^2 + \frac{c_2}{8} \Big( \big(\psi^* \tilde{\grad} \psi \big)^2 + \big(\psi \tilde{\grad} \psi^* \big)^2 \Big)  \Big\rbrace,
\end{align}
where $\tilde{\grad}$ denotes the gradient with respect to $\tilde{\Vec{x}}$ and $c_1 = \frac{2\hat{\beta}^2}{\hat{\delta}}$ and $c_2 = \frac{\hat{\beta}\hat{\mu}}{\hat{\delta}}$. The values of $c_1$ and $c_2$ in one, two and three dimensions are listed in TABLE \ref{tab: Microscopic coefficients}.
In the rescaled coordinates, to be explicit, the length is measured in units of $L_0 = \frac{1}{q_0} = \frac{v_F}{T_c} \ell_0$, where
\begin{equation}
    \ell_0 = \ell_0 \left( \frac{T}{T_c}, \frac{h}{T_c} \right) = \frac{1}{4 \pi} \sqrt{\frac{\hat{\delta}}{3 \hat{\beta}}} \frac{T_c}{T} \sqrt{\frac{\text{Re} \left[  \Psi ^{(4)}(z)\right]}{\text{Re} \left[  \Psi ^{(2)}(z)\right]}}
\end{equation}
is a dimensionless quantity that diverges as the tricritical point is approached. In the main text, we drop the tilde notation but still work in the rescaled model.
\begin{table}[H]
    \centering
    \begin{tabular}{c|c c c|c c}
         $d$ & $\hat{\beta}$ & $\hat{\delta}$ & $\hat{\mu}$ & $c_1$ & $c_2$ \\
         \hline
         \hline
         1 & 1 & $1/2$ & 4 & 4 & 8 \\
         \hline
         2 & $1/2$ & $3/16$ & 2 & $8/3$ & $16/3$ \\
         \hline
         3 & $1/3$ & $1/10$ & $4/3$ & $20/9$ & $40/9$
    \end{tabular}
    \caption{Numerical values of coefficients $c_1$ and $c_2$ in one, two, and three dimensions. The coefficients are computed using the microscopically derived values of the coefficients $\hat{\beta}, \hat{\delta}$, and $\hat{\mu}$.}
    \label{tab: Microscopic coefficients}
\end{table}

\section{Connection to the Swift-Hohenberg equation} \label{appendix: connection to SH}

The equation of motion corresponding to the Ginzburg-Landau model discussed in the main text reads
\begin{equation}
\begin{aligned}
\left( \alpha - 4 |\psi|^2 + 3 |\psi|^4 \right)  \psi + c_1 \laplacian{\psi} + \frac{c_1}{2} \nabla^4\psi + c_2 \Big\lbrace 
\psi |\grad{\psi}|^2  \\ - \divergence (|\psi|^2 \grad{\psi}) 
+ \frac{1}{4} \left[ \psi^* (\grad{\psi})^2 - \divergence (\psi^2 \grad{\psi^*}) \right] \Big\rbrace = 0.
\end{aligned}
\end{equation}

If we constrain $\psi$ to be real and rescale it by $\psi = u \left( \frac{c_1}{6} \right)^{1/4}$, we obtain the following equation for $u$
\begin{equation}\label{eq:SH}
-(1 + \laplacian)^2 u - \mu u + \nu u^3 - u^5 + \gamma \left( u (\grad{u})^2 + u^2 \laplacian u \right) = 0,
\end{equation}

where $\mu = \frac{2 \alpha}{c_1} - 1$, $\nu = 4 \sqrt{\frac{2}{3 c_1}}$ and $\gamma = \frac{5 c_2}{2 \sqrt{6 c_1}}$.
Note, that \eqref{eq:SH}  becomes the static cubic-quintic Swift-Hohenberg equation, in the limit $\gamma = 0$.

\section{Linearization} \label{appendix: linearization}

\begin{figure}[b]
    \centering
    \includegraphics[width=0.45\textwidth]{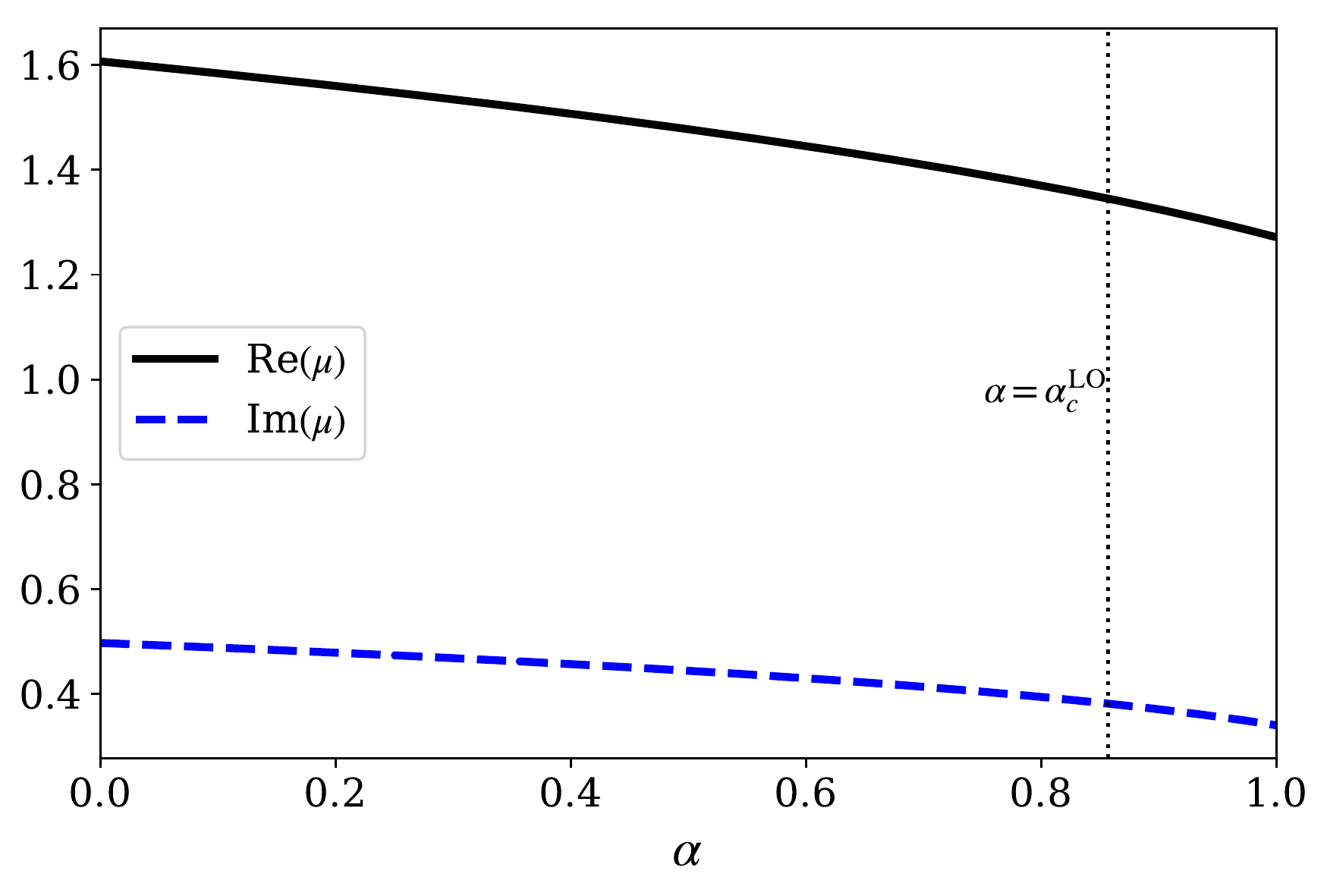}
    \caption{Plot of the inverse length scale $\mu$ for the system with changing $\alpha$ parameter. The inverse of the real part sets the decay length while the inverse of  imaginary part sets the oscillation length scale. Note that the transition to the LO state has been marked on the plot by the dotted line.} \label{fig: lengthscales}
\end{figure}

In this Appendix we will show the technical details as to how the linearized solutions presented in this paper for the long-range behavior of the field were found. 
 From our numerical solutions it follows that there is a class of solutions which can be described by a real field; thus in the analysis below we can
 restrict the field $\psi$ to be real. The resulting equation of motion reads
\begin{equation}
    \frac{dV}{d\psi} + c_1 \big( 2\laplacian{\psi} + \nabla^4 \psi \big) - \frac{5 c_2}{2} \big( \psi (\grad{\psi})^2 + \psi^2 \laplacian{\psi} \big) = 0, \label{eq: equation of motion}
\end{equation}
where $V(\psi) =  \alpha \psi^2 -2\psi^4 +\psi^6$ is the potential density. However, as we are interested in the behavior of the soliton far from it's  center, we write the field as,
\begin{equation}
\psi(r) = \psi_{\rm{U}} + \varepsilon(r),
\label{eq:eps}
\end{equation}
where we assume that the deviation $\varepsilon$ from the uniform ground state $\psi_{\rm{U}}$ has only radial dependence, is real, and is small. 
We can then proceed by considering the resulting equation of motion
for $\varepsilon$ by neglecting any terms in {\eqref{eq: equation of motion}} that are nonlinear in $\varepsilon$, as they will be negligible at long range.
This results in the linearized equation of motion
\begin{equation}
    c_1 \nabla^4 \varepsilon + 2\left(c_1-\frac{5c_2 \psi_{\rm{U}}^2}{4} \right) \laplacian{\varepsilon} + \left.\frac{d^2V}{d\psi^2}\right\rvert_{\psi= \psi_{\rm{U}}} \varepsilon = 0.
    \label{eq:lin_eom}
\end{equation}
Let us define the coefficients
\begin{align}
    a & = 1 - \frac{5c_2}{4c_1}\psi_{\rm{U}}^2, \\
    b & = \frac{1}{c_1}\left.\frac{d^2V}{d\psi^2}\right\rvert_{\psi= \psi_{\rm{U}}} = \frac{2}{c_1} \left( \alpha -12 \psi_{\rm{U}}^2 + 15\psi_{\rm{U}}^4 \right),
\end{align}
such that the linearized equation of motion reads
\begin{equation}
    \nabla^4 \varepsilon + 2a \laplacian{\varepsilon} +b \varepsilon = 0,
\end{equation}
which can be written, by introducing $\omega = \laplacian{\varepsilon}$, as a system of coupled differential equations
\begin{equation}
    (\laplacian{}-M)
    \begin{pmatrix}
    \omega \\ \varepsilon
    \end{pmatrix}
    = 0,
    \qquad
    M = \begin{pmatrix}
    -2a & -b \\ 1 & 0
    \end{pmatrix}.
\end{equation}
By linear transformation to the eigenbasis of the matrix $M$, the two equations decouple into
\begin{equation}
    \laplacian{\phi}-\mu^2 \phi = 0, \label{eq: decoupled Bessel equation}
\end{equation}
where $\mu^2 = -a \pm \ii \sqrt{b-a^2}$ are the two eigenvalues of the matrix $M$. We identify \eqref{eq: decoupled Bessel equation} as the modified Bessel equation, where for each eigenvalue of $M$, the solution in general is given as a superposition of the modified zeroth order Bessel functions $K_0$ and $I_0$. The Bessel function $I_0$ can be discarded directly by considering the asymptotic behavior at $r \to \infty$, and we obtain
{
\begin{equation}
    \phi_+  = C_+ K_0 (\mu r), \qquad
    \phi_-  = C_- K_0 (\mu^* r), 
\end{equation}
}
where $C_\pm$ are some complex constants, and where we have defined $\mu = \sqrt{ -a + \ii \sqrt{b-a^2} }$ with $\text{Re} (\mu)>0, \text{Im}(\mu)>0$, which is shown for the relevant parameter regime in \figref{fig: lengthscales}. 
The small deviation $\varepsilon$ is some superposition of $\phi_+$ and $\phi_-$. However, since $\varepsilon$ is real valued, we can use that
$K_0(z^*) = K_0(z)^*$ and consequently $\text{Re} \big[ K_0(z^*) \big] = \text{Re} \big[ K_0(z) \big]$,
which gives us
\begin{equation}\label{eq: linearisation}
    \varepsilon = \text{Re} \left[ C K_0(\mu r) \right],
\end{equation}
where $C$ is some complex-valued constant. 

\section{Long-distance interaction} \label{appendix: interaction}
 Next we follow the point-source approach \cite{Speight_vortex_forces} to determine the intersoliton forces. To that end, let
 us determine the point source $\rho$ that replicates the asymptotic field $\varepsilon$ given by \eqref{eq: linearisation}. The point source is defined by the equation
\begin{equation} \label{eq: definition of point source}
   c_1 \left( \nabla^4 \varepsilon + 2 a \laplacian \varepsilon + b \varepsilon \right) = \rho
\end{equation}
and can be found by considering the limit $|\Vec{x}|=r \to 0$, where $K_0 (\mu |\Vec{x}|) \to -\ln |\Vec{x}|$. Using that $\nabla^2 \ln |\Vec{x}|= 2\pi \delta (\Vec{x})$, we find
\begin{align}
    \nabla^2 K_0 (\mu |\Vec{x}|) & = \mu^2 K_0 (\mu |\Vec{x}|) - 2\pi \delta (\Vec{x}), \\
    \nabla^4 K_0 (\mu |\Vec{x}|) & = \mu^4 K_0 (\mu |\Vec{x}|) - 2\pi \Big(\mu^2 \delta (\Vec{x}) + \nabla^2 \delta (\Vec{x}) \Big),
\end{align}
where we have used that $(\laplacian - \mu^2) K_0(\mu |\Vec{x}|) = 0$ for $\Vec{x} \neq \Vec{0}$. Inserting these results into \eqref{eq: definition of point source} gives the point source
\begin{equation} \label{eq: formula point source}
    \rho = -2\pi c_1 \text{Re}\Big( C \big[ \nabla^2 \delta (\Vec{x}) - (\mu^*)^2 \delta (\Vec{x})\big] \Big).
\end{equation}

Having calculated the appropriate point source, we now consider two solitons, centered around $\Vec{x}_1$ and $\Vec{x}_2$, respectively, where the distance $|\Vec{x}_1-\Vec{x}_2|$ is large. To estimate the interaction between the two solitons, we assume that the total field $\psi$ is given by the superposition $\psi (\Vec{x}) = \psi_{\rm{U}} + s_1 (\Vec{x}) + s_2 (\Vec{x})$, where $s_j$ is the deviation from the ground state $\psi_{\rm{U}}$ for one soliton, centered around $\Vec{x}_j$. That is, $s_j (\Vec{x})$ approaches $\varepsilon_j (\Vec{x})$ far from its center. The main assumption here is that the field of the first soliton at the position of the second is the same as it would have been if there were no soliton there, and vice versa. This is quite a crude approximation since the center contributes significantly to the interaction energy.
The interaction energy $F_{\rm{int}}=F_{12}-F_1-F_2$ is derived by expanding to first order in the value of $s_1$ near $\Vec{x}_2$, and vice versa. After several integrations by parts we can get rid of the exact solitons $s_{1,2}$ in favor of their asymptotics $\varepsilon_{1,2}$, resulting in the interaction energy
\begin{equation}
    F_{\rm{int}} = -\int_{\mathbb{R}^2} \rho_1 \varepsilon_2 dx dy = -\int_{\mathbb{R}^2} \rho_2 \varepsilon_1 dx dy.
\end{equation}
By using the derived expression for the asymptotic in \eqref{eq: linearisation} and the point source in \eqref{eq: formula point source} we find
\begin{equation}
    F_{\rm{int}} = - 2 \pi c_1 \sqrt{b - a^2} \text{Im}\Big( C^{(1)} C^{(2)} K_0( \mu|\Vec{x}_1-\Vec{x}_2| ) \Big),
\end{equation}
where $C^{(1,2)}$ are the constants associated with the two soliton asymptotics respectively. The oscillatory nature of $K_0(\mu|\Vec{x}_1-~\Vec{x}_2| )$ implies that the solitons will be weakly bound at large distances.

\bibliography{references.bib} 

\end{document}